\documentstyle[12pt,epsf]{article}
\def \insps#1#2#3#4#5#6#7 {
  \vspace{-#4}
  \epsfxsize #3
  \epsfysize #4
  \begin{figure}[#6]
    \centerline{\epsfbox{#2}}
    \caption{#7}
    \label{#1}
  \end{figure}}

\begin{document}
\begin{center}
{\bf \LARGE HIJING Model Prediction On Squeeze-Out of Particles in
Nucleus-Nucleus Interactions at Super High Energies}
\end{center}

\large

\begin{center}
{\bf A.S. Galoyan, V.V. Uzhinskii}
\end{center}

\begin{center}
Joint Institute for Nuclear Research, Dubna, Russia
\end{center}

\begin{abstract}
\noindent Elliptic flow of hard partons ($P > 5$ GeV), squeeze-out of
soft partons ($P \leq 5$ GeV) and produced particles are predicted
in the framework of the HIJING  model. They are caused due to jet
quenching and heterogeneity of the interaction region.
\end{abstract}

Collective flows of produced particles and nuclear fragments in
nucleus-nucleus interactions have been studied for the last 15 years
(see the latest reviews in \cite{Review1, Review2, Review3}). Five 
years ago, the so-called radial flow of nuclear fragments was 
discovered in central collisions of gold nuclei at intermediate 
energies (see \cite{Radial1} -- \cite{Radial5}). Recently, it was 
recognized that the elliptic flow of produced particles changes its 
sign at energies 5 -- 10 GeV/nucleon \cite{Elliptic}. It is astonished 
that the collective flows are saved in interactions of lead nuclei at 
the energy of 158 GeV/nucleon \cite{PbPb}. They may be observed 
at RHIC and LHC.

Now, it is commonly accepted that the hot and compressed nuclear matter 
is created in a central participant region (a reaction zone) formed by 
the overlap of projectile and target nuclei during the early stage of 
the nucleus-nucleus collisions at intermediate energies. This 
compression leads to the flow in the reaction plane where the majority 
of spectators are concentrated. The direction perpendicular to the 
reaction plane is the only direction where the nuclear matter might 
escape during the whole collision time without being hindered by either 
the target or projectile nucleus \cite{Kampert}. As a consequence, this 
leads to a jet-like emission pattern, also referred to as "out-of-plane
squeeze-out" \cite{Stoker}. Squeeze-out was first predicted by
a hydrodynamic model \cite{Stoker}. In terms of microscopic models, the
collective motions are caused by (re)scattering. Providing that the
mean free path of the particles is smaller than the dimensions of the
spectators, the (re)scattering leads to exactly the same picture.

It is hard to expect that this scenario will be realized at RHIC.
Nevertheless, the direct HIJING model calculations presented in the 
Fig.  1 show a clear preferential emission of mid-rapidity particles in 
the direction perpendicular to the reaction plane (at $\varphi \sim 
90^o$ and $\varphi \sim 270^o$). In the figure, we plot the azimuthal 
distributions $dN/d\varphi$ of charged and neutral particles with $\mid
\eta \mid \leq 5.4$ around the beam axis in $Au+Au$-interactions at
$\sqrt{s_{NN}}=200$ GeV.  $\varphi$ is the azimuthal angle of a
particle relative to the azimuth of the reaction plane.  Since HIJING
model \cite{HIJING} does not take into account the interactions of
produced particles with spectator matter, it is clear that the effect
is caused by the jet quenching \cite{Quench} and heterogeneity of the
interaction region.

In the HIJING model \cite{HIJING}, energy losses of hard partons (jet
quenching) are imitated by production of soft collinear gluons. For
example, the partons associated with {\it i}-th wounded nucleon of 
nucleus A having radius-vector $\vec r_i =(x_i, y_i)$ in the plane 
perpendicular to the reaction plane (in the impact parameter plane)
(see fig. 2) can interact with either the wounded nucleons of A nucleus
or with the wounded nucleons of B nucleus. Interaction takes place
if the impact parameter in the collision of the parton of {\it i}-th 
nucleon having transverse momentum $\vec P_T$ with {\it j}-th nucleons 
of B nucleus is less than 1 fm \footnote{In the subroutine QUENCH of 
the HIJING program the quantities $\pm b/2$ are omitted. There were 
another mistakes in the subroutine. The corrected version of the HIJING model
see at home-page http://nt3.phys.columbia.edu/people/molnard/OSCAR/}, 

$$ 
b' ~=~ \mid [\vec P_T \times 
(\vec r_j\prime - \vec b/2 -(\vec r_i + \vec b/2))]\mid / \mid \vec P_T 
\mid \leq 1~~(fm).  $$ The energy loss of the parton is determined as 
$$
\Delta E ~=~C \mid [\vec P_T \bullet (\vec r_j\prime - \vec b/2 -(\vec
r_i + \vec b/2))]\mid / \mid \vec P_T \mid ,
$$
$C=<dE/dz> ~=~ 2$ GeV/fm is the average energy loss per unit of length.
This energy loss is ascribed to a new gluon associated with {\it j}-th
nucleon of B nucleus collinear to the parton.

Since the density of the wounded nucleons in the impact parameter plane
at non-zero impact parameter is heterogeneous (it has y-size large
than x-size, see Fig. 2), a condition for the flow effects 
appears.  For example, a parton disposed in the center of the region
and flown with $\varphi \sim 90^o$, or $270^o$ to the X-axis will loose
in the average more energy than a parton flown with $\varphi \sim 0^o$,
or $180^o$. As it was said above, the energy losses are imitated by
soft gluon production. So, the soft gluons will  have to be
preferential emitted perpendicular to the impact parameter vector.
The hard gluons will  have to do the same, but to be emitted alonge
the impact parameter vector. The calculations presented in the Fig. 3
confirm these expectations.

As seen, there must be elliptic flow of the partons with $P_g >  5$
GeV with amplitude $v_2 \sim 0.003 \div  0.006$, and squeeze-out of the
partons with $P_g \leq 5$ GeV with amplitude $v_2 \sim -0.006 \div
-0.019$ in $Au ~+~ Au$ interactions at $\sqrt{s_{NN}} ~=~ 200$ GeV
according to the HIJING model.

Strings are formed in the HIJING model from the partons. Observable
hadrons appear at the string fragmentation. Thus, the squeeze-out
presented in the calculations of the Fig. 1 is caused by the processes.

According to the Fig. 1, the amplitude of the squeeze-out flow 
increases with increasing the particle momentum. For example, at
$b ~< ~ 9$ fm $v_2 ~=~ -0.001$, $-0.002$, $-0.008$ for the particles
with $P \leq 2$ GeV, $2 < P \leq 5$ Gev, and $P > 5$ GeV, respectively.
The amplitude vanishes in peripheral interactions.
The maximum flow can be at intermediate values of the impact parameter
(see the bottom row of the Fig.  1).

It should be note, that the accounting of the energy losses in the
HIJING program is implemented too simple, and asymmetric with respect
to A and B nuclei. In principle, there must be a normal cascade
production of the gluons. Though, the heterogeneity of the wounded
nucleon distribution in the impact parameter plane, what can be
calculated as in the HIJING model, as in the quisi-eikonal approach
\cite{QUSI} is a real source of flow effects in nucleus-nucleus
interactions. We believe that the flows will be discovered in current
RHIC experiments.

The authors thank the members of the CMS collaboration for
useful discussion of the problem.

\insps{fig1}{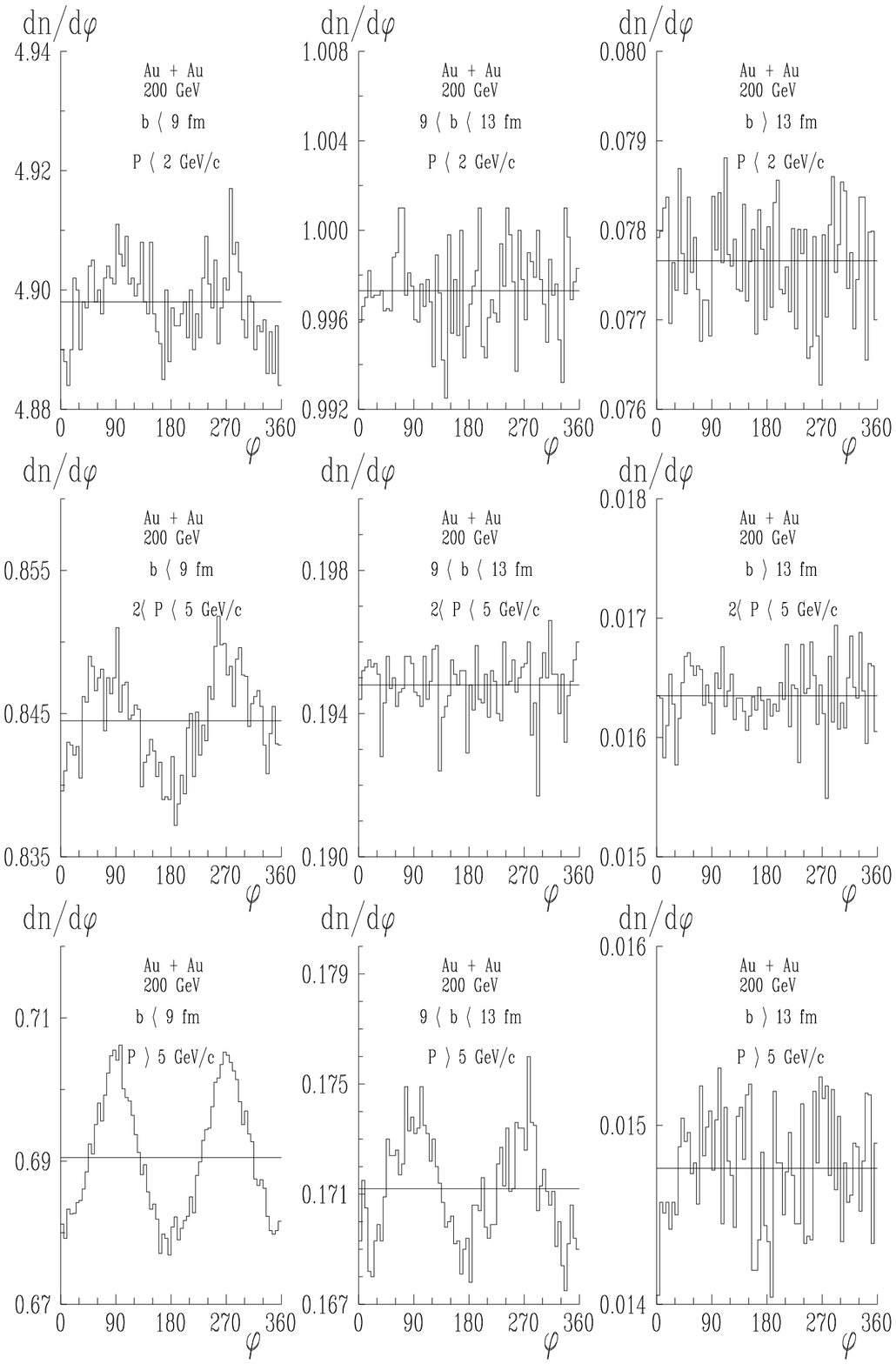}{6in}{9in}{0in}{here}{Azimuthal distributions of
particles in Au+Au interactions}

\insps{fig2}{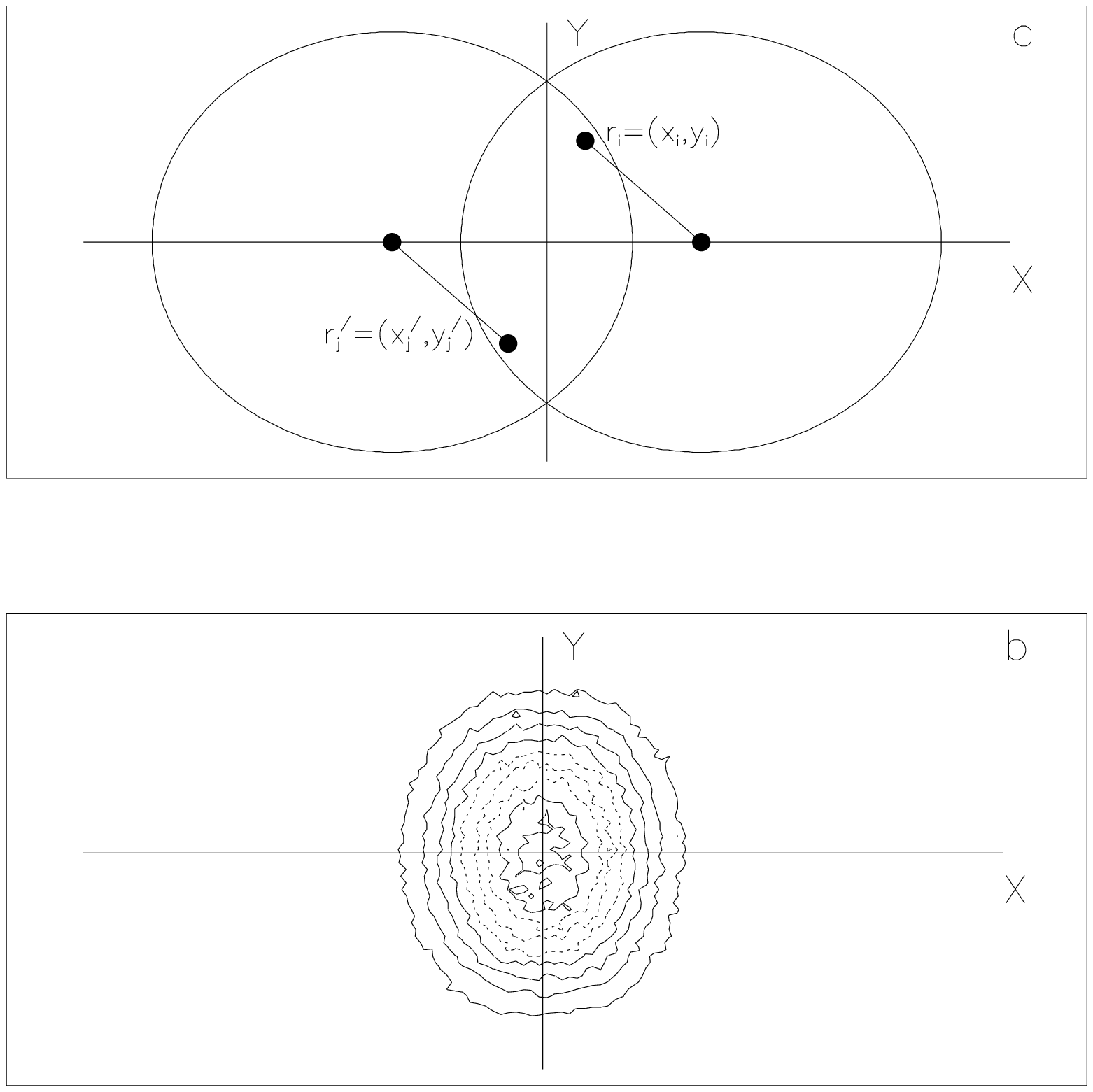}{6in}{9in}{0in}{here}{a) Geometry of a collision in
the impact parameter plane. b) Wounded nucleon density in the reaction
zone.}

\insps{fig3}{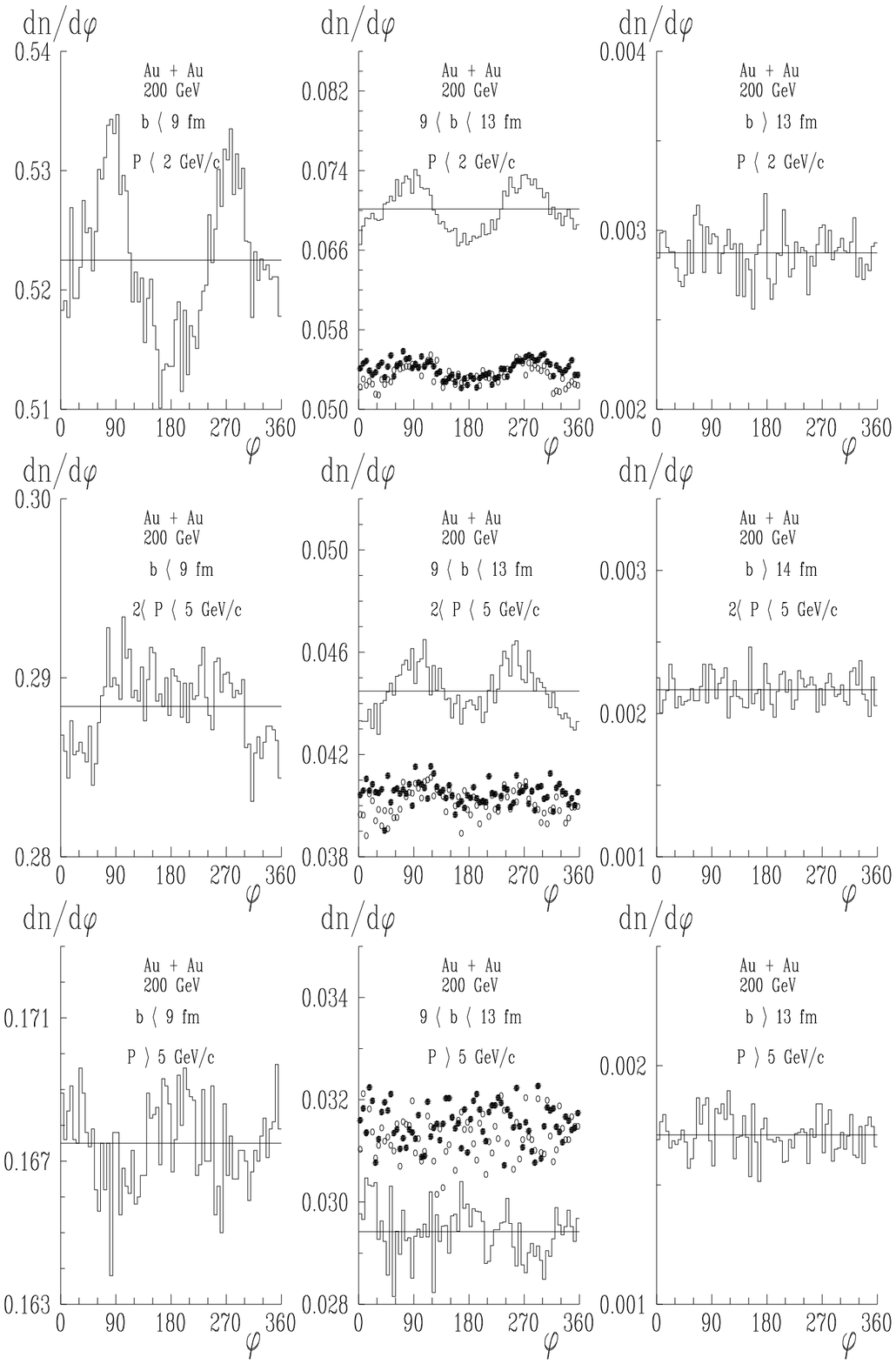}{6in}{9in}{0in}{here}{Azimuthal distributions of
partons in Au+Au interactions. Open and closed points present the
calculations with quenching of jet associated only with A, or B
nucleus, respectively.}

\end{document}